\documentclass[12pt]{article}

\usepackage{amsmath,amsfonts,amssymb,latexsym}

\newtheorem{theorem}{Theorem}[section]

\newenvironment{proof}[1][Proof]{\textsc{#1.} }{\ \rule{0.5em}{0.5em}}
\numberwithin{equation}{section}
\def\be{\begin{equation}}
\def\ee{\end{equation}}
\def\bq{\begin{eqnarray}}
\def\eq{\end{eqnarray}}
\def\beq{\begin{eqnarray*}}
\def\eeq{\end{eqnarray*}}

\begin{document}
\begin{titlepage}
\begin{flushright}
\end{flushright}

\vspace{1cm}

\begin{center}
{\huge Singular Isotropic Cosmologies and Bel-Robinson Energy\footnote{to appear in the Proceedings of the Albert Einstein Century International Conference, Paris, France, July 18-22, 2005.}}

\vspace{1cm}

{\large Spiros Cotsakis$\dagger$ and Ifigeneia
Klaoudatou$\ddagger$}\\

\vspace{0.5cm}

{\normalsize {\em Research Group of Geometry, Dynamical Systems
and Cosmology}}\\ {\normalsize {\em Department of Information and
Communication Systems Engineering}}\\ {\normalsize {\em University
of the Aegean}}\\ {\normalsize {\em Karlovassi 83 200, Samos,
Greece}}\\ {\normalsize {\em E-mails:} $\dagger$
\texttt{skot@aegean.gr}, \texttt{$\ddagger$ iklaoud@aegean.gr}}
\end{center}

\vspace{0.7cm}
\begin{abstract}
We consider the problem of the nature and possible types of
spacetime singularities that can form during the evolution of
\emph{FRW} universes in general relativity. We show that by using,
in addition to the Hubble expansion rate and the scale factor, the
Bel-Robinson energy of these universes we can consistently
distinguish between the possible different types of singularities
and arrive at a complete classification of the singularities that
can occur in the isotropic case. We also use the Bel-Robinson
energy to prove that known behaviours of exact flat isotropic
universes with given singularities are generic in the sense that
they hold true in every type of spatial geometry.
\end{abstract}


\end{titlepage}
\section{Introduction}
There has recently been a resurgence of interest in the old
question of the existence and nature of spacetime singularities,
especially in a cosmological context. Traditionally one is
interested in formulating criteria, of a geometric nature, to test
under what circumstances such singularities, in the form of
geodesic incompleteness, will be formed during the evolution in
general relativity and in other metric theories of gravity, cf.
\cite{he73}. These are usually translated into sufficient
conditions to be satisfied by the matter fields present and are
very plausible. On the other hand, one can  formulate equally
plausible and generic geometric criteria for the long time
existence of geodesically complete, generic spacetimes in general
relativity and also other theories of gravity, cf. \cite{chc02}.
Such criteria assume a globally hyperbolic spacetime in the
so-called sliced form (cf. \cite{c04}), require the space gradient
of the lapse function as well as the extrinsic curvature not to
grow without bound, and are also very plausible.

The lesson to be drawn from this state of affairs is that in
general relativity (and also in other metric gravity theories),
singular and complete spacetimes are equally generic in a sense.
The real question is what do we mean when we say that a
relativistic model develops a spacetime singularity during its
evolution. In other words, what are the possible spacetime
singularities which are allowed in gravity theories? These may be
singularities in the form of geodesic incompleteness developing in
the course of the evolution but also others which are more subtle
and which are also dynamical and present themselves, for instance,
in some higher derivatives of the metric functions spoiling the
smoothness of global solutions to the field equations.

Obviously a recognition and complete analysis of such a program
cannot be accomplished in the short run and requires complete
examination of a number of different factors controlling the
resulting behaviour. For instance, one needs to have control of
each possible behaviour of the different families of relativistic
geometry coupled to matter fields in general relativity and other
theories of gravity to charter the possible singularity formation.
In this sense, unraveling the nature and kinds of possible
singularities in the simplest kinds of geometry becomes equally
important as examining this problem in the most general solutions
to the Einstein equations. In fact, an examination  of the
literature reveals that the more general the spacetime geometry
considered (and thus the more complex the system of equations to
be examined) the simplest is the type of singularities allowed to
be examined. By starting with a simple cosmological spacetime we
allow all possible types of singularities to come to surface and
be analyzed.

In \cite{ck05} we derived necessary conditions for the
existence of finite time singularities in globally and regularly hyperbolic
isotropic universes, and provided first evidence for their nature based entirely on the behaviour of the Hubble
parameter $H=\dot{a}/a$.  This result may be summarized as
follows:
\begin{theorem}\label{1}
Necessary conditions for the existence of finite time singularities in
globally hyperbolic, regularly hyperbolic \textsc{FRW} universes are:
\begin{description}
\item[$S_{1}$] For each finite $t$, H is non-integrable on $[t_1,t]$,
or
\item[$S_{2}$] H blows up in a finite future time, or
\item[$S_{3}$] H is defined and integrable for only a finite
proper time interval.
\end{description}
\end{theorem}

Condition $S_{1}$ may describe different types of singularities.
For instance, it describes  a big bang type of singularity, when
$H$ blows up at $t_{1}$ since then it is not integrable on any
interval of the form $[t_{1},t]$,  $t>t_{1}$ (regular
hyperbolicity is violated in this case, but the scale factor is
bounded from above). However, under  $S_{1}$ we can have other types of
singularities: Since $H(\tau)$ is integrable on an interval
$[t_1,t]$, if $H(\tau)$ is defined on $[t_1,t]$,  continuous on
$(t_1,t)$ and the limits $\lim_{\tau\rightarrow t_1^+}H(\tau)$ and
$\lim_{\tau\rightarrow t^-}H(\tau)$ exist, the violation of
\emph{any} of these conditions leads to a singularity that is not
of the big bang type discussed previously.

Condition $S_{2}$ describes
a future blow up singularity and condition $S_{3}$ may lead to a
sudden singularity (where $H$ remains finite), but for this to be
a genuine type of singularity, in the sense of geodesic
incompleteness, one needs to demonstrate that the metric is
non-extendible to a larger interval.

Note that these three
conditions are not overlapping, for example $S_{1}$ is not implied by
$S_{2}$ for if $H$ blows up at some finite time $t_s$ after $t_1$,
then it may still be integrable on $[t_1,t]$, $t_1<t<t_s$.

There are many examples in the literature of singularities belonging to
the types predicted by the above theorem; they were
analyzed in \cite{ck05} in detail. They usually describe flat isotropic
universes with various components of matter. For example, those models
containing phantom dark energy \cite{gonzales}, or a perfect fluid and a scalar
field \cite{mel}, exhibit future finite time blow up singularities and therefore
fall in the $S_{2}$ category. The standard big bang singularities as well as the
sudden singularity of \cite{ba04} fall in the $S_{1}$ category. Other sudden
singularities (which do not have a blow up of $H$ at $t=0$, see for example
\cite{noj1}) and the inflation model of \cite{borde} both fall in the $S_{3}$ category.

\section{Classification}
As discussed in \cite{ck05}, Theorem (\ref{1}) describes possible
time singularities that are met in $\textsc{FRW}$ universes
having a Hubble parameter that behaves like $S_{1}$ or $S_{2}$ or
$S_{3}$. Although such a classification is a first step to clearly
distinguish between the various types of singularities that can
occur in such universes, it does not bring out some of the
essential features of the dynamics that  differ from singularity
to singularity. For instance, condition $S_{2}$  includes both a
collapse singularity, where $a\rightarrow 0$ as $t\rightarrow
t_{s}$, \emph{and} a blow up singularity where $a\rightarrow
\infty$ as $t\rightarrow t_{s}$. Such a degeneration is unwanted
in any classification  of the singularities that can occur in
the model universes in question.

It is therefore necessary to refine this classification by
considering also the behavior of the scale factor. Another aspect
of the problem that must be taken into account is the relative
behaviour of the various  matter components as we approach the
time singularities.

In what follows we present a summary of our most recent work on
the classification of possible singularities which occur in
isotropic model universes. Details will be published elsewhere,
cf. \cite{ck2}. Our classification is based on the introduction of
an invariant geometric quantity, the Bel-Robinson energy, which
takes into account precisely those features of the problem, related  to the
matter contribution, in which models still differ
near the time singularity while having similar behaviours of $a$
and $H$. In this way, we arrive at a complete classification of
the possible cosmological singularities in the isotropic case.

The Bel-Robinson energy is a kind of energy of the gravitational
field \emph{projected} in a sense to a slice in spacetime. It is
used in \cite{ck93}, \cite{chm} to prove global existence results
in the case of an asymptotically flat and cosmological spacetimes
respectively, and is defined as follows. Consider a sliced
spacetime with metric
\begin{equation}
^{(n+1)}g\equiv -N^{2}(\theta ^{0})^{2}+g_{ij}\;\theta ^{i}\theta ^{j},\quad
\theta ^{0}=dt,\quad \theta ^{i}\equiv dx^{i}+\beta ^{i}dt,  \label{2.1}
\end{equation}
where $N=N(t,x^{i})$ is the lapse function and $\beta^{i}(t,x^{j})$ is the
shift function, and the $2$-covariant spatial electric
and magnetic tensors
\beq
E_{ij}&=&R^{0}_{i0j},\\
D_{ij}&=&\frac{1}{4}\eta_{ihk}\eta_{jlm}R^{hklm},\\
H_{ij}&=&\frac{1}{2}N^{-1}\eta_{ihk}R^{hk}_{0j},\\
B_{ji}&=&\frac{1}{2}N^{-1}\eta_{ihk}R^{hk}_{0j},
\eeq
where
$\eta_{ijk}$ is the volume element of the space metric $\bar g$.
The \emph{Bel-Robinson energy} is given by \be
\mathcal{B}(t)=\frac{1}{2}\int_{\mathcal{M}_t}\left(|E|^{2}+|D|^{2}+
|B|^{2}+|H|^{2}\right)d\mu_{\bar{g}_t}, \ee where by
$|X|^{2}=g^{ij}g^{kl}X_{ik}X_{jl}$ we denote the spatial norm of
the $2$-covariant tensor $X$. In the following, we exclusively use
an $\textsc{FRW}$ universe filled with various forms of matter
with metric given by
\be ds^2=-dt^2+a^2(t )d\sigma ^2,
\ee
where
$d\sigma ^2$ denotes the usual time-independent metric on the
3-slices of constant curvature $k$. For this spacetime, we find that the norms
of the magnetic parts, $|H|, |B|$, are identically zero while $|E|$ and
$|D|$, the norms of the electric parts, reduce to the forms
\be
|E|^{2}=3\left(\ddot{a}/{a}\right)^{2} \quad\textrm{and}\quad
|D|^{2}=3\left(\left(\dot{a}/{a}\right)^{2}+k/{a^{2}}\right)^{2}.
\ee
Therefore the Bel-Robinson energy becomes
\be \mathcal{B}(t)=\frac{C}{2}\left(|E|^{2}+|D|^{2}\right),
\ee
where $C$ is the constant volume of (or \emph{in} in the case of a
non-closed space) the 3-dimensional slice at time $t$.

We can now classify the possible types of singularities that are
formed in an $\textsc{FRW}$ geometry during its cosmic evolution and
enumerate the possible types that result from the different
combinations of the three main functions in the problem, namely,
the scale factor $a$, the Hubble expansion rate $H$ and  the Bel
Robinson energy $\mathcal{B}.$ If we
suppose that the model has a finite time singularity at $t=t_s$, then
the possible behaviours of the functions in the triplet $\left(H,a,(|E|,|D|)\right)$
in accordance with Theorem (\ref{1}) are as follows:
\begin{description}
\item [$S_{1}$] $H$ non-integrable on $[t_{1},t]$ for every $t>t_{1}$

\item [$S_{2}$] $H\rightarrow\infty$  at $t_{s}>t_{1}$

\item [$S_{3}$] $H$ otherwise pathological
\end{description}

\begin{description}
\item [$N_{1}$] $a\rightarrow 0$

\item [$N_{2}$] $a\rightarrow a_{s}\neq 0$

\item [$N_{3}$] $a\rightarrow \infty$
\end{description}

\begin{description}
\item [$B_{1}$] $|E|\rightarrow\infty,\, |D|\rightarrow \infty$

\item [$B_{2}$] $|E|<\infty,\, |D|\rightarrow \infty $

\item [$B_{3}$] $|E|\rightarrow\infty,\, |D|< \infty $

\item [$B_{4}$] $|E|<\infty,\, |D|< \infty $.
\end{description}

There are a few types that cannot occur. For instance, we cannot have an
$(S_{2},N_{2},B_{3})$ singularity because that would imply having
$a<\infty$ ($N_{2}$) and $H\rightarrow\infty$ $(S_{2})$, while
$3\left( (\dot{a}/{a})^{2}+k/a^{2}\right)^{2}<\infty$ $(B_{3})$,
at $t_{s}$ which is impossible since $|D|^{2}\rightarrow\infty$ at $t_{s}$
($k$ arbitrary). A complete list of impossible singularities is given by $(S_{i},N_{j},B_{l})$, where the indices in the case of a
$k=0,\pm1$ universe take the values $i=1,2$, $j=1,2,3$, $l=3,4$,  whereas in the case of a $k=-1$ universe the indices take the values $i=1,2$, $j=2,3$, $l=3,4$ (here by $S_{1}$ we denote for simplicity only the big bang case in the  $S_{1}$ category). We thus see that some singularities which are impossible for a flat or a closed universe become possible for a hyperbolic universe. Consider for example the triplet $(S_{2},N_{1},B_{3})$ which means having
 $H\rightarrow\infty$, $a\rightarrow 0$ and
$$|D|^2=3\left(\left(\dot{a}/{a}\right)^{2}+k/{a^{2}}\right)^{2}<\infty$$
at $t_{s}$. This behaviour is valid only for some cases of a hyperbolic universe.

All other types of finite time singularities can in
principle be formed during the evolution of \textsc{FRW}, matter-filled models,
in general relativity or other metric theories of gravity.

It is interesting to note that all the standard dust or radiation-filled big-bang
singularities fall under the \emph{strongest} singularity type, namely, the type
$(S_{1},N_{1},B_{1})$. For example, in a flat universe filled
with dust, at $t=0$ we have
\bq
a(t)&\propto& t^{2/3}\rightarrow 0, \quad (N_{1}),\\
H&\propto& t^{-1}\rightarrow\infty, \quad (S_{1}),\\
|E|^{2}&=&3/4H^{4}\rightarrow\infty,\quad
|D|^{2}=3H^{4}\rightarrow\infty, \quad (B_{1}).
\eq
Note that the classification is organized in such  a way that the character of the
singularities becomes milder as the indices of $S$, $N$ and
$B$ increase. Milder singularities are thus met as one proceeds down the list.

In fluid-filled models, the various behaviours of the Bel-Robinson energy density can be related to four
conditions imposed on the density and pressure of the cosmological fluid:
\begin{description}
\item [$B_{1}$] $\Leftrightarrow$ $\mu\rightarrow \infty$ and $|\mu+3p|\rightarrow\infty$

\item [$B_{2}$] $\Leftrightarrow$ $\mu\rightarrow \infty$ and $|\mu+3p|<\infty$

\item [$B_{3}$] $\Leftrightarrow$ $\mu<\infty$ and $|\mu+3p|\rightarrow\infty$
$\Leftrightarrow$ $\mu<\infty$ and $|p|\rightarrow\infty$

\item [$B_{4}$] $\Leftrightarrow$ $\mu< \infty$ and $|\mu+3p|<\infty$
$\Leftrightarrow$ $\mu<\infty$ and $|p|<\infty$.
\end{description}
Of course we can translate these conditions to asymptotic behaviours
 in terms of $a,H$, depending on
the value of $k$, for example,
\begin{enumerate}
\item If $k=0$, $\mu<\infty$ $\Rightarrow$ $H^{2}<\infty$, $a$ arbitrary

\item If $k=1$, $\mu<\infty$ $\Rightarrow$ $H^{2}<\infty$ and $a\neq 0$

\item If $k=-1$, $\mu<\infty$ $\Rightarrow$ $H^{2}-1/a^{2}<\infty$.
\end{enumerate}
As an example, we consider the sudden singularity introduced in \cite{ba04}.
This has a finite $a$ (condition $N_{2}$), finite $H$ (condition $S_{3}$),
finite $\mu$ but a divergent $p$ (condition $B_{3}$)
at $t_{s}$. As another example, consider the  flat \textsc{FRW} model
containing dust and a scalar field studied in
\cite{mel}. The scale factor collapses at both
an initial (big bang) and a final time (big crunch). The Hubble
parameter and $\ddot{a}/a$  both blow up at the times of the big bang and  big crunch
 (cf. \cite{ck05}) leading to an $(S_{1},N_{1},B_{1})$ big bang singularity and
an $(S_{2},N_{1},B_{1})$ big crunch singularity, respectively.

\section{Generic results and examples}
In this Section we provide necessary and sufficient
conditions for the occurrence of some of the singularities
introduced above. These conditions are motivated from studies of cosmological models described by exact solutions
in the recent literature. By exact solutions we mean those in which
all arbitrary constants have been given fixed values. We expect the proofs of these results to be all quite
straightforward, for we have now already identified  the type of singularity that
we are looking for in accordance with our classification.
Proving such results \emph{without} this knowledge would have been a problem
of quite a  different order.

The usefulness of the results proved below lies in that they answer the question of whether
the behaviours met in  known models described by exact solutions
(which as a rule have a flat spatial
metric ($k=0$)) continue to valid in universes having all possible values of $k$ described
by solutions which are more general than exact in the sense that some or all of the arbitrary constants  still remain arbitrary. See \cite{ck2} for more results of this type.

\begin{theorem} \label{2}
Necessary and sufficient conditions for an $(S_{2},N_{3},B_{1})$
singularity occurring at the finite future time $t_{s}$ in an isotropic universe filled with a fluid with  equation of state
$p=w\mu$, are that $w<-1$ and $|p|\rightarrow\infty$ at $t_{s}$.
\end{theorem}
\begin{proof}

 Substituting the equation of state $p=w\mu$ in the
continuity equation $\dot{\mu}+3H(\mu+p)=0$, we have
\be
\label{phantom}
\mu\propto a^{-3(w+1)},
\ee
and so if $w<-1$ and $p$ blows up at $t_{s}$, $a$ also blows up at
$t_{s}$. Since
\be
H^{2}=\frac{\mu}{3}-\frac{k}{a^{2}},\quad
|D|^{2}=\frac{\mu^{2}}{3},
\quad
|E|^{2}=\frac{1}{12}\mu^{2}(1+3w)^{2},
\ee
we conclude that at $t_{s}$, $H$, $a$, $|D|$ and $|E|$ are divergent.

 Conversely, assuming an $(S_{2}, N_{3}, B_{1})$ singularity at $t_{s}$ in
 an \textsc{FRW} universe with the equation of state
$p=w\mu$, we have from the $(B_{1})$ hypothesis that $\mu\rightarrow\infty$
at $t_{s}$ and so $p$ also blows up at $t_{s}$. Since $a$ is divergent as
well, we see from (\ref{phantom}) that $w<-1$.
\end{proof}

As an example, consider an exact solution which describes a flat, isotropic
phantom dark energy filled universe, studied in \cite{gonzales} given by
\be
\alpha={[{\alpha _{0}}^{3(1+w)/2}+\frac {3(1+w)\sqrt
{A}}{2}(t-t_{0})]}^{\frac {2}{3(1+w)}},
\ee
where $A$ is a constant. From
(\ref{phantom}) we see that the scale factor, and consequently
$H$, blows up at the finite time $$t_{s}=t_{0}+\frac{2}{3\sqrt {A}
(|w|-1)\alpha_{0}^{3(|w|-1)/2}}.$$ Then it follows that
$|E|^{2}=\frac{3}{4}H^{4}(1+3w)^{2}$ and $|D|^{2}=3H^{4}$ also blow up at $t_{s}$.
Therefore in this model the finite time singularity is of type
$(S_{2},N_{3},B_{1})$.

The following result says that the strongest big bang type singularities that can occur in
universes with a massless scalar field are due to the kinetic term.
\begin{theorem} \label{6}
A necessary and sufficient condition for an $(S_{1},N_{1},B_{1})$
singularity at $t_{1}$ in an isotropic universe with a massless scalar field is that
$\dot{\phi}\rightarrow\infty$ at $t_{1}$.
\end{theorem}

\begin{proof}
From the continuity equation,
$
\ddot{\phi}+3H\dot{\phi}=0,
$
we have
$\dot{\phi}\propto a^{-3}$, and if $\dot{\phi}\rightarrow\infty$
then $a\rightarrow 0$. Since
\be
H^{2}=\frac{\mu}{3}-\frac{k}{a^{2}}\rightarrow\infty
\ee
$H$ becomes unbounded at $t_{1}$. In addition, since
\be
|D|^{2}=\frac{\mu^{2}}{3}=\frac{\dot{\phi}^{4}}{12}\rightarrow\infty,
\ee
and
\be
|E|^{2}=\frac{1}{12}(\mu+3p)^{2}=\frac{\dot{\phi}^{4}}{3}\rightarrow\infty,
\ee
at $t_{1}$, both $|D|$ and $|E|$ diverge there.

Conversely, assuming an $(S_{1},N_{1},B_{1})$
singularity at $t_{1}$, we have (from $B_{1}$) that $\mu\rightarrow\infty$
and so $\dot{\phi}^{2}\rightarrow\infty$ at $t_{1}$.
\end{proof}

As an example of the above behaviour we can use the exact solution
for a flat isotropic universe given in \cite{fo}. The solution is
given by $$H=\frac{1}{3t}, \quad
\phi=\pm\sqrt{\frac{2}{3}}\ln\frac{t}{c}.$$ Since $a\propto
t^{1/3}$ we have that at $t=0$, $a\rightarrow 0$ ($N_{1}$),
$H=\frac{1}{3t}\rightarrow \infty$ ($S_{1}$),
$|E|^{2}=\frac{2}{9}\dot{\phi}^{4}$ and
$|D|^{2}=3H^{4}\rightarrow\infty$ ($B_{1}$). As it follows from
\cite{fo}, this exact solution represents the asymptotic behaviour
of a scalar field model if
\be
\lim_{\phi\rightarrow\pm\infty}e^{-\sqrt{6}|\phi|}V(\phi)=0.
\ee

\section{Discussion}
We have reviewed some of the main results of our recent work \cite{ck2} regarding the classification of singularities in
isotropic universes. The possible behaviours of the three functions, $H$, $a$ and $\mathcal{B}$ exhaust the types of singularities that
are possible in an isotropic universe. We have used known
exact solutions as examples to illustrate the possible singularities that
can be accommodated in the categories included in the classification. We have also shown
that the Bel-Robinson energy can be used invariable as a tool to test
the nature of singularities described by given exact solutions and decide
whether such behaviour is generic and independent of  the spatial geometry.

It would be interesting to further investigate how this
classification will change when we consider other types of model
universes, matter fields different from the fluid ones we considered in
this paper, or other metric theories of gravity.

We note that the "crudest" singularity type  $(S_{1},N_{1},B_{1})$, seems
to accommodate all standard big bang singularities known. However, the
precise analytical nature of the other new singularity types presented here,
although certainly possible, remains a mystery. It is only through a combination of analytic and geometrical techniques that more light will be shed in such interesting questions of principle.

\section*{Acknowledgements}
This work was supported by the joint Greek Ministry of Education and European Union  research
grants `Pythagoras' and `Heraclitus' and this support is gratefully acknowledged.

\end{document}